# Quantum Computation—The Ultimate Frontier

**Chris Adami and Jonathan P. Dowling**

*Quantum Computing Technologies Group*
*Jet Propulsion Laboratory MS 126-347, California Institute of Technology,*
*Pasadena, CA 91109*

The discovery of an algorithm for factoring which runs in polynomial time on a quantum computer has given rise to a concerted effort to understand the principles, advantages, and limitations of quantum computing. At the same time, many different quantum systems are being explored for their suitability to serve as a physical substrate for the quantum computer of the future. I discuss some of the theoretical foundations of quantum computer science, including algorithms and error correction, and present a few physical systems that have shown promise as a quantum computing platform. Finally, we discuss a spin-off of the quantum computing revolution: quantum technologies.

## Introduction

The theory of computation has undergone a radical renewal over the last seven years, thanks to Peter Shor's discovery (Shor 1994) that numbers can be factored in *polynomial* (rather than "sub-exponential") time, on a quantum computer. The significance of this discovery can perhaps only be appreciated if we compare the time it takes to factor a 400-digit integer on a classical computer with the currently best algorithm (the number-sieve), to what could conceivably be achieved with Shor's algorithm running on a quantum computer (see Fig. 1)

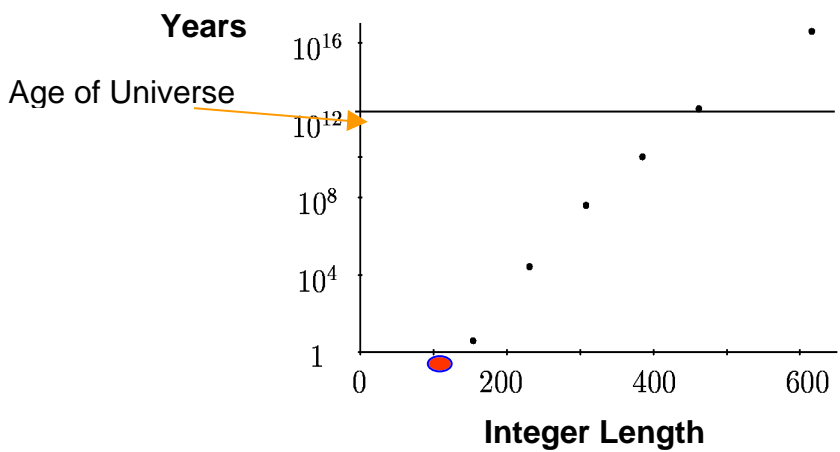

FIG. 1: Time to factor integers of increasing length (in digits). Current best is indicated by red dot.

While the classical algorithm will be busy for as long as the universe is old (on the order of billions of years), a quantum algorithm, because the time scales only as the square as the length of the integer, would finish this task within a day, possibly in seconds. Such a difference is of course eye-opening, in particular because the difficulty of factoring is the key ingredient in the most common encryption mechanisms used today (those based on RSA). As I will point out further below, it may be less staggering if it would turn out that a polynomial classical algorithm exists after all, but in that case the existence of a polynomial quantum factoring algorithm would almost certainly have led us to that discovery.

The "power of the quantum" is clearly not immediately obvious to the computational scientist. On the contrary, common wisdom taught that quantum mechanics imposes a fundamental limit to the speed and accuracy of computations. Indeed, Moore's Law, interpreted as predicting the number of electronic components on a computing chip, predicts that the average number of elementary objects participating in a computation approaches unity around the year 2012. This number implies that single electrons or single atoms would be involved in such computations, the behavior of which, however, need to be described by quantum mechanics, and are subject to the uncertainty principle. Uncertainty in computation, it is perhaps superfluous to say, is not welcome. A standard examination of this uncertainty shows that it is due to the uncontrollability of quantum degrees of freedom that interact with their environment, as well as to the general concept of "vacuum fluctuations". These factors appear to spell the end of Moore's Law.

## Entanglement and the Qubit

On the other hand, uncertainty in quantum mechanics is not absolutely inevitable. Consider, for example, Schrödinger's equation. Does it not predict the evolution of a quantum state in a purely deterministic manner? Isn't this evolution unitary, that is, probability conserving? The answer is: "Yes indeed", and this is precisely the foundation on which quantum computation is built. However, we have to be very careful. Unitary evolution applies to wave functions, not to individual objects. In classical mechanics, we do not usually make a distinction between the objects and their mathematical description, but this difference is crucial in quantum mechanics, due to the property of entanglement. Entanglement is what happens to the state of two quantum mechanical objects that, before their interaction, each had a wave function of their own. After their interaction, *one* wave function describes the two-object system, even if one of the objects is sent to the other side of the known universe. While, in our classical minds, we anticipate that each of the objects has a state independent of the other (in particular if one of them is at Pluto) quantum mechanics (in theory and experiment) tells us that we are wrong. This invisible bond between the objects is what makes quantum mechanics interesting, and quantum computation possible. Let us then move to a more formal description of a quantum computer's foundations.

Computation is still carried out on bits, but they are now *quantum bits* (qubits for short). Many substrates have been suggested (and used) to embody quantum bits, but perhaps the most well known (if not practically the most useful) is the spin-1/2 system. Electrons, for example, have half-integer spins, and can exist in two-different helicity states (the projection of their spin along their direction of motion): parallel or anti-parallel, plus or minus, up or down, zero or one. A perfect two-state system, it so seems. But qubits quickly reveal that they are more complex than that. While a qubit has these two basic states (we will denote them using Dirac's "ket" notation as |0> and |1>, respectively), a qubit can in fact "be" in any superposition of these states. Thus, its wave function can be

$$|\ \rangle = \ |0\rangle + \ |1\rangle, \qquad (1)$$

where    and    are complex numbers. We put the word "be" in quotation marks in the sentence above because, in fact, the concept of reality becomes complicated in quantum mechanics. The reason for this is simple to see. In order to determine an object's state we have to interact with it. In general, interacting with it implies entangling your (the observer's) wave function with that object. After this, you (the observer) do not have a state any more: you cannot (from the point of view of the mathematics of the object's as well as the observer's wave function) be thought of independently of each other anymore, you are forever entangled. It is usual, in quantum mechanics, to attempt to determine a quantum system's state by repeated measurements, and to determine the wave function *before* measurement by constructing one that agrees best with the observed outcomes. However, this requires us to be able to reconstruct the system whose state we are interested in over and over again. (In quantum mechanics, this is called a "known" or "prepared" state.) If we can do this, why do we need to measure the state?. Thus we are left with the conclusion that an "unknown" (meaning arbitrary) quantum state of the form given in Equation (1) is problematic from an epistemological point of view: We have to accept that its quantum reality is strictly decoupled from our physical reality that is reflected in experiments.

The mathematics of quantum mechanics (such as Eq. 1) are perfectly unambiguous and its predictions agree perfectly well with experiment. Then, the schism between quantum and physical realism need not bother us as long as we keep aware that "seeing" an object take on a particular state (observing its physical reality) may not reveal to us the wave function of this object (its quantum reality). Let us now return to quantum states such as Eq. (1). Two such systems can interact via a unitary operation (the "entanglement operation") $U_{ent}$ in the following manner:

$$U_{ent} |\ \rangle |0\rangle = \ |0\rangle|0\rangle + \ |1\rangle|1\rangle \qquad (2)$$

In the above equation, | > could for example stand for an unknown quantum system, and |0> a known one that we would perhaps use as a measurement device (this would perhaps be "you"). As we warned above, the resulting state is not a product of two wave functions (such as, say, the wave function | >| >). The operator $U_{ent}$ is an important primitive in quantum computation. Its power becomes obvious if we apply it (or more precisely, its *N*-qubit extension) to *N* qubits that have been prepared in a product state:

$$U_{ent} \underbrace{|0\rangle|0\rangle|0\rangle|0\rangle|0\rangle\ldots\ldots|0\rangle|0\rangle}_{N \text{ times}} \quad |00000\ldots00\rangle + |00000\ldots01\rangle + |00000\ldots10\rangle +$$

$$+ \ldots + |11111\ldots00\rangle + |11111\ldots01\rangle + |11111\ldots10\rangle + |11111\ldots11\rangle \ (2^N \text{ terms}). \quad (3)$$

Thus, *N* qubits can, using the entanglement operation, be brought into a superposition of $2^N$ terms, each of which can be thought of as encoding a particular classical state. (Any product state can be viewed as a "classical" state.) The power of quantum computation is that the $2^N$ terms in superposition (3) can be operated on simultaneously with a unitary quantum operator $U$ (this is known as *quantum parallelism*). Of course, you need to make sure that the entanglement extends to all these qubits, and these qubits *only*. Should a few qubits become entangled with (3) inadvertently (which means, should you not be aware that this happened), the wonderful properties of the entangled states will have turned from boon to bane: Instead of a coherent superposition you will be handed an uncertain mixture of states, useless for computation. In a manner of speaking, then, all quantum computations ought to be performed in "total darkness" (since photons are perfect qubits on their own), and at zero temperature. Even "looking" at your end result could prove to be hazardous if special precautions are not taken..

These problems, seemingly insurmountable at first, have nevertheless been solved (again by Shor, 1995 and others) by transferring techniques from classical error correction into quantum mechanics. Today, error-correcting codes for arbitrarily large quantum registers can be constructed, and techniques have been devised to tolerate small errors in the operations performed on them (see, e.g., Shor 1996, Preskill 1998).

## Building a Quantum Computer

Quantum computing requires the creation, manipulation, and detection, of entangled qubits. Each of these requirements is important, and poses different challenges depending on the physical representation of a qubit. We have seen earlier that quantum mechanical spin is the most obvious embodiment of a qubit. However, the half-integral spin of electrons, for example, is very difficult to control. Nuclear spins are used for computation in NMR (nuclear magnetic resonance) quantum computers. To design such a computer, a molecule with an atom for each qubit (see Fig. 2) has to be designed in such a manner that each qubit can be addressed independently by suitable bursts of microwaves that can flip individual nuclear spins.

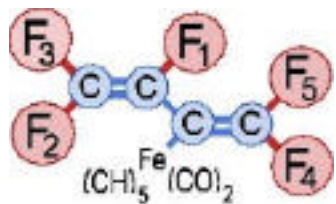

FIG. 2: Custom-designed Fluorine molecule used as a quantum computer to solve a simple problem (courtesy IBM Research).

While the technology is important to demonstrate quantum logic, it is difficult to scale and perhaps does not represent the physical implementation of choice for the future. Other implementations are the quantum states of atoms. The ground state and an excited state of suitably prepared atoms can be designated the "zero" and "one" states, and laser pulses allow the manipulation of the state of each one, while results can be shifted from one atom to an adjacent one (Fig. 3).

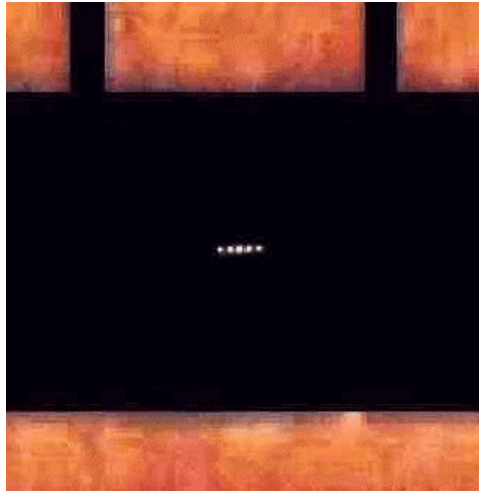

FIG. 3: Photograph of five beryllium ions in a lithographically fabricated RF trap. The separation between ions is approximately 10 microns (courtesy D. Wineland, NIST).

There is no shortage of attempts to embody quantum bits in physical quantum states. Magnetic flux tubes can carry quantized units of flux, and be used in silicon-based systems as qubits. So can Cooper pairs (pairs of electrons in a superconducting state) in quantum dots. In that case, the charge state (which is quantized) plays the role of the qubit. In solid-state quantum computing, phosphorus dopants in a silicon array act as qubits, and are manipulated and accessed using metallic gates on the surface of the chip together with external ac and dc magnetic fields (Fig. 4).

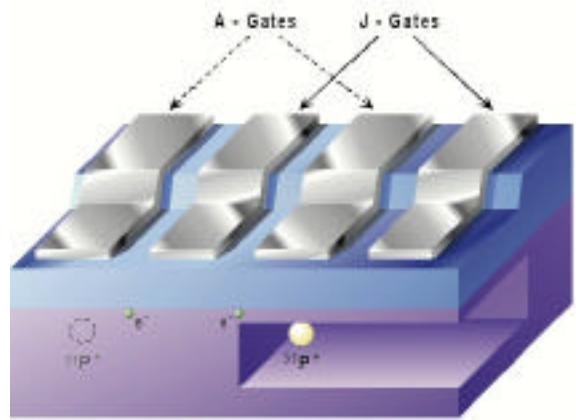

FIG. 4: Silicon-based quantum computing architecture (courtesy University of New South Wales).

Silicon-based implementation carry the promise of scalability, but problems of decoherence (the loss of control of the qubits involved) is still a major problem. Other implementations use the polarization of photons as qubits, in all-optical implementations of quantum computers. While these usually require non-linear optical elements, progress has been made to replace the non-linear elements by the clever use of projective measurements and quantum teleportation techniques (Knill et al., 2001).

While work on putative quantum computing platforms is proceeding slowly, an offspring of the quantum computing revolution is now developing quickly and garnering more and more attention: the idea of quantum technology and quantum metrology. That control of quantum degrees of freedom should result in ultra-precise measurement devices should not have come as a surprise if we had always been aware of Landauer's Principle (Landauer, 1961), namely that computation and measurement are fundamentally the same processes, that they are just different faces of the same coin. We could then have predicted that the quantum methods that sped up computation to "superclassical" levels would give rise to measurement devices with superclassical resolution.

## Quantum Technology: The 2$^{nd}$ Quantum Revolution

The history of the future of physics is undergoing a monumental shift that we like to call the *Second Quantum Revolution.* The *First Quantum Revolution* began at the end of the last century—and the current revolution comes at the dawn of the new millennium. Physics and technology for the next 100 years will be dominated by the technological advances associated with this new revolution in *Quantum Technology* (Milburn 97). The Department of Defense is in a prime position to utilize the scientific breakthroughs that will come from harnessing novel quantum effects, such as quantum coherence, entanglement, and nonlocal correlations. These are the very same physical resources underlying the exploding fields of *Quantum Information Technology, Quantum Optics, Quantum Atomics, and Coherent Quantum Electronics.* There is a synergy between all these fields, in that an advance in one can often be turned into an advantage in the others.

The *First Quantum Revolution* occurred at the last turn of the century, arising out of theoretical attempts to explain experiments on blackbody radiation. From that theory arose the fundamental idea of wave-particle duality—in particular the idea that matter particles sometimes behaved like waves, and that light waves sometimes acted like particles. This simple idea underlies nearly all of the scientific and technological breakthroughs associated with this *First Quantum Revolution*. Once you realize just how an electron acts like a wave, you now can understand the periodic table, chemical interactions, and electronic wavefunctions that underpin the electronic semiconductor physics. The latter technology drives the computer-chip industry and the Information Age. On the other hand, the realization that a light wave must be treated as a particle gives to us the understanding we need to explain the photoelectric effect for constructing solar cells and photocopying machines. The concept of the photon is just what we need to understand the laser. By the end of this century, this first revolution of quantum mechanics has evolved into many of the core technologies underpinning modern society. However, there is *a Second Quantum Revolution* coming—which will be responsible for most of the key physical technological advances for the 21st Century.

The hallmark of this *Second Quantum Revolution* is the realization that we humans are no longer passive observers of the quantum world that Nature has given us. In *the First Quantum Revolution*, we used quantum mechanics to understand what already existed. We could explain the periodic table, but not make new atoms. We could explain how metals and semiconductors behaved, but not do much to manipulate that behavior. The difference between science and technology is the ability to engineer your surroundings to your own ends, and not just explain them. In the *Second Quantum Revolution*, we are now actively employing quantum mechanics to alter the quantum face of our physical world—developing a *quantum* technology. We are transforming Nature into highly unnatural quantum states of our own design, for our own purpose. For example, in addition to explaining the periodic table, we can make new artificial atoms—quantum dots and excitons—that we can engineer to have electronic and optical properties of our own choosing. We can create states of quantum coherent or entangled matter and energy that hitherto existed nowhere else in the Universe. These new man-made quantum states have novel properties of sensitivity and nonlocal activity that have wide application to the development of computers, communications systems, sensors and compact metrological devices—all areas of intense DoD interest. Thus, although quantum mechanics as a science has matured completely, quantum engineering as a *technology* is now emerging on its own right. It is just a matter of being in the right place at the right time to take full advantage of these exciting new developments.